\documentclass[aps,twocolumn,superscriptaddress,showpacs]{revtex4}

\usepackage{natbib,graphicx,bm,url}

\begin{document}
\def\bl{{\lambda_{avg}}}

\author{Jacob R. Wintersmith}
\affiliation{Department of Physics, Harvey Mudd College, Claremont, CA 91711. }

\author{Lu Zou}
\affiliation{Department of Physics, Kent State University, Kent, OH 44242.}
.
\author{Andrew J. Bernoff}
\affiliation{Department of Mathematics, Harvey Mudd College, Claremont, CA 91711.}

\author{James C. Alexander}
\affiliation{Department of Mathematics, Case Western Reserve University, Cleveland, OH 44106.}

\author{J. Adin Mann, Jr.}
\affiliation{Department of Chemical Engineering, Case Western Reserve University, Cleveland, OH 44106.}

\author{Edgar E. Kooijman}
\affiliation{Department of Physics, Kent State University, Kent, OH 44242.}

\author{Elizabeth K. Mann}
\affiliation{Department of Physics, Kent State University, Kent, OH 44242.}

\title{Determination of Inter-Phase Line Tension in Langmuir Films}
\date{\today}

\begin{abstract}
A  Langmuir film  is a molecularly thin film  on the surface of a fluid; we study the evolution of a Langmuir film with two co-existing fluid phases driven by an inter-phase line tension and damped by the viscous drag of the underlying subfluid. Experimentally, we study an 8CB Langmuir film via  digitally-imaged Brewster Angle Microscopy (BAM) in a  four-roll mill setup which applies a transient strain and images the response. When a compact domain is stretched by the imposed strain,  it first assumes a bola shape with two tear-drop shaped reservoirs connected by a thin tether which then slowly relaxes to a circular domain which minimizes the interfacial energy of the system. 
We process the digital images of the experiment to extract the domain 
shapes. We then use one of these shapes as an initial condition 
for the numerical solution of a boundary-integral model of the 
underlying hydrodynamics and  compare the subsequent 
images of the experiment  to the numerical simulation.
The numerical evolutions first verify that our hydrodynamical model can reproduce the observed dynamics. They also allow us to deduce  the magnitude of the line tension in the system, often to within 1\%. We find line tensions in the range of 200-600 pN; we hypothesize that this variation is due to differences in the layer depths of the 8CB fluid phases.

\end{abstract}
\maketitle

Line tension, the two dimensional analog of surface tension, is the free energy per unit length
associated with the boundary between two phases  on a surface. In this paper we explore a method for measuring the inter-phase line tension in Langmuir layers, the quasi-two-dimensional surface layers of polymers, lipids or liquid crystals that exist at gas-liquid and liquid-liquid interfaces.
Langmuir layers often separate into multiple domains signaling the coexistence of different phases
\citep{Adamson98}. The boundaries of such domains are curved, yielding a line force per unit length normal to the phase boundary and tangent to the surface containing the Langmuir layer  with a magnitude that is the product of the  line tension and the curvature of the interphase boundary.

Attempts to measure the line tension in various systems have multiplied over recent
years.  One motivation is to better understand the forces which govern the shape and influence the
function of biological membranes; cell membranes consist of a
mixture of cholesterol, lipids, and proteins that can form domains
with various structures and functions. Model membranes, including supported bilayers \citep{Stottrup04}, vesicles \citep{Baumgart03} and Langmuir monolayers \citep{Adamson98} show macroscopic phase separation, with geometry driven by line tension.

Line tension between fluid Langmuir phases has most often been
measured by watching the relaxation of stretched domains toward an energy-minimizing 
circular shape. The relaxation of large perturbations, such as 
bola-shaped domains (two teardrop-shaped resevoirs tethered together with a line
of nearly constant thickness) have been modelled only heuristically;
models to extract line tension \citep{BenvegnuM92, Mann95, Lauger96} 
approximated the bola shape as two perfectly round discs connected by an
infinitesimally thin tether, which is far from the
true form. The dynamics of  linearized perturbations of circular domains are better understood
\citep{StoneM95, Mann95, JFM}, but these perturbations are difficult to measure accurately in the small amplitude limit where they obtain validity. Due to these problems, the error bounds  of previous line tension measurements have been no better than $\pm 20\%$.

Our group recently  developed a manageable model \citep{JFM} of the
experimentally observed relaxation dynamics of two  fluid
phases within a Langmuir film. The model is both analytically
tractable and allows an efficient, accurate and stable numerical solution via a boundary-integral technique.

In this article we directly compare the numerical
results of our model to experimental results on a Langmuir layer with two fluid phases corresponding to different multilayer thicknesses, and test both the validity of the model and the precision of the line tension measurements resulting from that
comparison. We expect this to set the stage for further accurate and
precise studies of line tension as a function of temperature,
composition, and other variables. 

\section{Experimental}

We conduct our measurements on Langmuir films
comprised of 4$^\prime$-8-alkyl[1,1$^\prime$-biphenyl]-4-carbonitrile (\emph{8CB}) deposited on a subfluid of pure water. The \emph{8CB} exists as a smectic
liquid crystal with stacked molecular bilayers on
top a simple monolayer at the water surface \citep{deMul94,deMul98, Lauger96}. Consequently, multiple phases each consisting of a different odd number of layers (i.e. monolayers, trilayers, etc.)  can simultaneously exist within the film.

Relaxation in Langmuir layers is driven by intermolecular forces between the surface molecules and also between the layer  and the subfluid. In some systems, electrostatic forces in the Langmuir layer (primarily dipole-dipole repulsion) drive interesting pattern formation such as circle-to-dogbone transitions and labyrinth formation \citep{DeKokerM93, McConnell91}.
We choose  the \emph{8CB} multilayer system considered here because the electrostatic effects
are probably negligible, in that a symmetric bilayer is added at
each step. No jump in surface potential, which determines the
effective dipole moment density, is observed after the triple layer.

Consequently, in this system the intermolecular forces are well-modeled as a line tension at
phase boundaries, which
causes the film to coalesce into spatially-distinct
phase-domains.
Any domain strained into a non-circular shape will relax to the
energy-minimizing circular configuration, driven by the inter-phase line
tension.

The \emph{8CB} forms a smectic phase at the water surface which behaves like a two-dimensional fluid. The surface viscosity can be estimated from the bulk viscosities of the
smectic phase; it is less than $100$ times the viscosity of water \citep{Chen91,Mukai97}, so that for domains we consider with  thickness less than $100$ nm, the surface viscosity is negligible as long
as the domain size is $ \gg 10 \mu$m. 

The \emph{8CB} (Sigma-Aldrich, 98\% pure) is further purified by chromatography. 
We dissolve \emph{8CB} in hexane (Fischer, Optima grade)
spreading solution, which is deposited on the surface of water (PureLab+ system, 
and passes the shake test) in a clean trough (mini-trough, KSV). After
deposition, the hexane evaporates, leaving an \emph{8CB} layer on the water
surface. The trough has a pair of movable barriers to change the
water surface area available to the \emph{8CB} film and thereby
control the surface pressure. At room temperature, surface pressures
$\sim \!\! 6.5$ mN/m produce a stable coexistence of a tri-layer
over the entire surface interspersed with thicker domains
\citep{deMul98, deMul94}. We image the Langmuir layer
using a homemade Brewster Angle Microscope (BAM) \citep{Henon91, Honig91, Zou04}, which produces grey-scale images showing more thickly-stacked
domains in brighter shades against the dark, thinly-stacked
background.

\begin{figure}
\includegraphics[width=30mm]{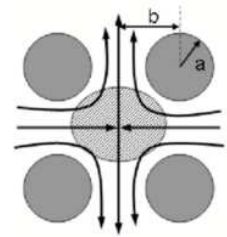}
\caption{Schematic of the four-roll-mill flow profile.
The geometry of the rollers, $a = 6.6$ mm and $b = 10.5$ mm, produces
maximum homogeneity in the extension rate \citep{Higdon93, Fuller97, Kooijman00},. The shaded ellipse shows the area illuminated by the
laser beam.\label{fig:4RM}}
\end{figure}

We stretch the domains by shearing the subfluid and then use the BAM
to observe the subsequent relaxation, which is recorded on a
computer at 30 frames per second. To shear the subfluid we use a
4-roll mill \citep{Higdon93, Fuller97, Kooijman00} controlled by a
stepper motor. The rolls are made of black Delrin which is
hydrophilic and has no measurable effect on surface pressure. We adjust the
water level to be exactly the same height as the upper edges of the
rolls in order to minimize the distortion of the fluid surface
resulting from contact with the rolls. As shown in Figure
\ref{fig:4RM}, the 4-roll mill provides symmetric shear forces
about a central stagnation point on the surface. This allows us to
stretch a domain located at the stagnation point without imparting a
net velocity to the domain and moving it out of the BAM's field of
view. A controlled air stream maneuvers a domain into proper
location at the stagnation point. Once the domain is in position, we
activate the 4-roll mill, and the domain stretches out, assuming the
characteristic bola shape . Generally, we run the 4-roll mill at
speeds of $\sim \!\! 0.2$ revolutions per second for about 5
seconds. In our experiments, the Reynolds number \citep{Lagnado90} 
of the flow during shearing is $\sim \!\! 16$. Because of the
inertia in the subfluid, the domain continues stretching for several seconds  after the mill has been stopped.

\section{Hydrodynamics}

Our model \citep{JFM} describes the dynamics of a
Langmuir layer consisting of two phases: an isolated phase-domain,
$\Omega$, of finite area surrounded by a second Langmuir phase,
$\Omega^C$, which extends infinitely in the horizontal direction.
The Langmuir layer is modeled as a flat, two-dimensional fluid.
We assume that the subfluid is infinitely deep. Both the
Langmuir layer and the subfluid are assumed to be incompressible on the timescale of the relaxation experiments. Thus, the Langmuir domain $\Omega$ will  have a fixed area,  $A_*$.

For the Langmuir layer, the incompressibility condition in relaxation driven
by line tension alone corresponds to a Gibbs elasticity $\kappa _{g}\gg
\lambda /\sqrt{A_*}$, where $\lambda$ is the inter-phase line tension \cite{Alexander06}.
For the experimentally accessible range we estimate that 
$\sqrt{A_*}>10\mu m $ and conservatively choose an upper limit on the line tension of 
$\lambda <1nN$,  which yields an upper bound on the Gibbs elasticity, 
$\kappa _{g}\gg 10^{-4}N$, for significant compressibility.
Thus, almost any Langmuir layer liquid and many gasses
will act as incompressible fluids in line-tension driven relaxation
processes (see \cite{Alexander06} and references therein).  
In the special case of  multilayers considered here, it is
additionally conceivable that the number of layers might change during the
relaxation process, leading to an area change. In practice, the thickness of the layers 
and the area of the domains were observed to be 
constant for the duration of the experiments  reported here.

As we discuss above, dimensional analysis  indicates that for \emph{8CB}  the energy
dissipated by viscous shearing within the Langmuir layer is much
less than the amount dissipated by viscous shearing of the subfluid;
we therefore model the Langmuir layer as inviscid. Furthermore, the subfluid can be treated in
the Stokesian limit, where its inertia is negligible.

We non-dimensionalize the dynamics in terms of a
characteristic length, time, and mass
\begin{equation} \label{eq:ND}
L_* = \sqrt{A_*}, \;\; T_* = \frac{\eta' A_*}{\lambda}, \;\; M_* = \eta' L_* T_*,
\end{equation}
respectively; here $\lambda$ is the inter-phase line tension and
$\eta'$ is the subfluid viscosity. Essentially,  the relaxation of
the domain $\Omega$ is driven forward by the line tension between
phases, and slowed by the viscosity of the subfluid.

The model ultimately yields an equation of motion for the boundary
curve, $\vec{\Gamma}$, separating $\Omega$ and $\Omega^C$. As the boundary is isotropic, it suffices to determine the normal velocity, ${\cal U}$, to specify the domain's evolution. We \mbox{obtain \citep{JFM}}
\begin{equation}
\label{eq:EOM}
{\cal U} = \left( \frac{\partial \Psi}{\partial s}  \right) \hat{n}
\end{equation}
where $\hat{n}$ is the outward unit normal vector to $\vec{\Gamma}$, $s$ is the arclength
measured in a right-handed sense, and $\Psi(s)$ is the velocity streamfunction restricted to the boundary of the domain. This is computed as a boundary integral, 
\begin{equation}
\Psi(s) =  - \frac{1}{2\pi} \oint_{\vec{\Gamma}} 
 \kappa(s') \left  [  \hat{t}(s') \cdot \hat{Q}(s,s')  \right ] ~ ds'        \mbox{ ,}
\end{equation}
where $\kappa$ is the curvature, $\hat{t}$ is the unit tangent vector, and
$\hat{Q}(s,s')$ is a unit vector pointing from $\vec{\Gamma}(s)$ to $\vec{\Gamma}(s')$. A derivation and discussion  of this formulation is given in \citep{JFM}.

We implement a numerical solution in \emph{MATLAB}. The problem is extremely stiff numerically; explicit integration methods are very susceptible to high-wavenumber instabilities. This can be ameliorated by operator splitting, following the ideas of \citet{HouLS94}. While such a
splitting in not immediately apparent in the formulation above, the
formulation in \citet{LubenskyG96} and \citet{HeinigHF04} can be
used to show that the high-frequency modes of $\vec{\Gamma}$ are asymptotically governed by a much simpler evolution law, namely motion by mean curvature.

As in \citep{HouLS94}, using an intrinsic description of the boundary  allows an accurate
implicit solution for the high-wavenumber modes, avoiding numerical
instabilities. We represent the boundary with an equal-arclength
discretization.  Derivatives are computed pseudo-spectrally
\citep{Gottlieb77,Trefethen00}, and the boundary integral is
computed using the 16-panel closed Newton-Cotes formula which
guarantees high-order spatial accuracy.  It is straightforward to
solve the evolution by mean curvature implicitly and to high
accuracy \citep{HouLS94}. We proceed by using Strang splitting
\citep{Strang68} with the mean-curvature step implemented implicitly
and  the boundary integral velocity minus
the mean curvature velocity computed explicitly.

Numerically, we observe a slow drift of the grid which forces us to regularly correct the arclength discretization --- this is done using spectral interpolation and Newton-Raphson iteration. Also, it is necessary to filter the
 highest-frequency modes of $\vec{\Gamma}$ (whose numerical
accuracy is poor due to the discretization anyway);  we
convolve the spectrum with a smooth filter and retain roughly
two-thirds the spectrum. Details of the numerical implementation 
are available  in \citep{Pugh06}.

\section{Finding the Boundary Curves}
To analyze a set of experimental
photographs we must first determine the location of the boundary
curve in each one.  A grey-scale photograph is a map from each pixel $(x,y)$ to the brightness of the image at that location, $B(x,y)$. The edge of the domain is located in the
region of rapid transition from black to white, where $|| \nabla B ||$ is large. We compute $\nabla B$ and $
|| \nabla B ||$, using code developed by Fisher et al  \citep{Canny}. We
execute a curve-tracing algorithm that ``walks'' around the edge of
the domain, staying in the thin region where $|| \nabla B ||$ is
large. As the algorithm traverses the boundary it marks points, which
we subsequently use as a discrete representation of the boundary
curve.

The placement of the edge can be quickly verified visually. We also have a quantitative
check at our disposal. The domain area is conserved; if the edge is placed too far to the
outside or inside then as the perimeter of the domain decreases during
relaxation the computed area of the domain decreases or
increases, respectively. This relationship allows us to calculate the
(average) distance by which the edge is displaced in the normal
direction. In most data sets we see no correlation between perimeter
length and domain area. When this effect is seen, the implied displacement  is never
more than two pixels, and we can move the edge
in the normal direction to correct for the displacement (this was done in data sets B and E reported below).

The greatest obstacle to determining the precise location of the
edge is diffraction, which blurs the edge and produces a
bright ring of constructive interference. Although the average normal displacement of the curve is very small, diffraction may cause the edge to  be off by several pixels locally (this problem affects both human visual perception and computer algorithms).

\begin{figure}
\begin{tabular}{c}
\includegraphics[width= 80mm]{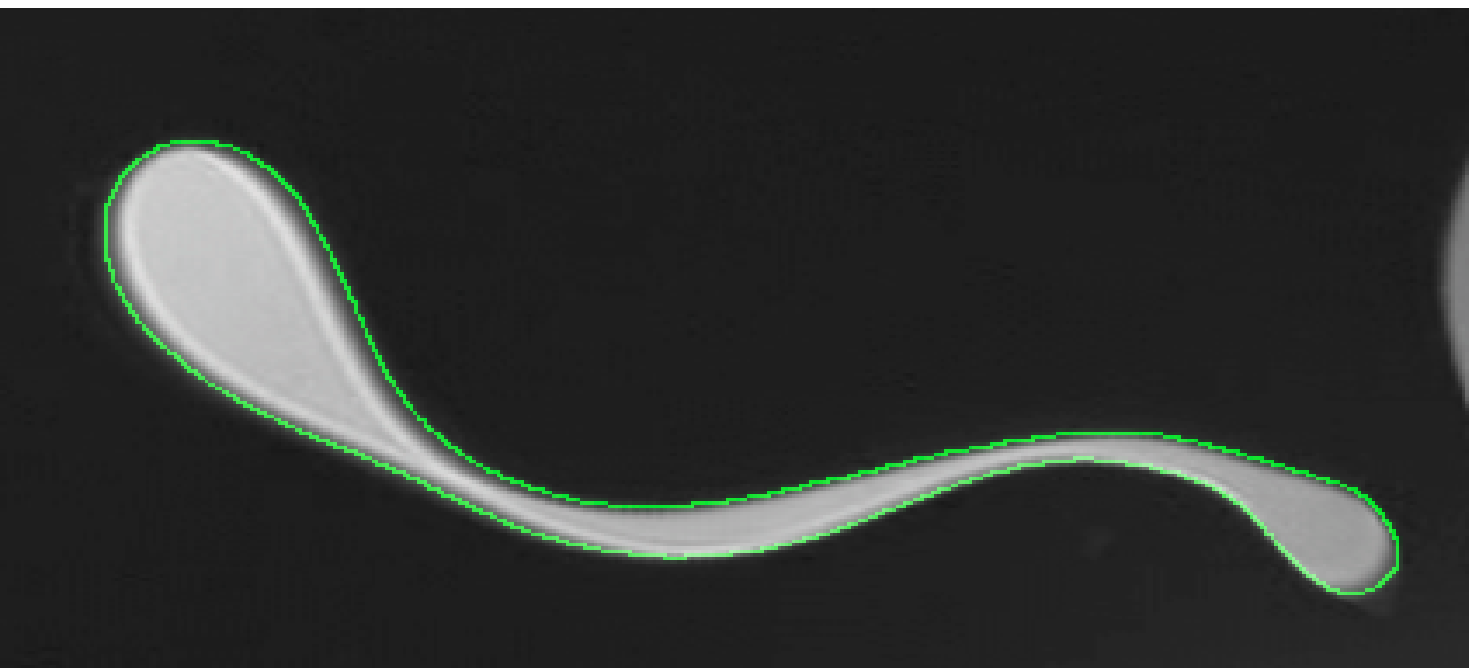}  \\
\includegraphics[width= 80mm]{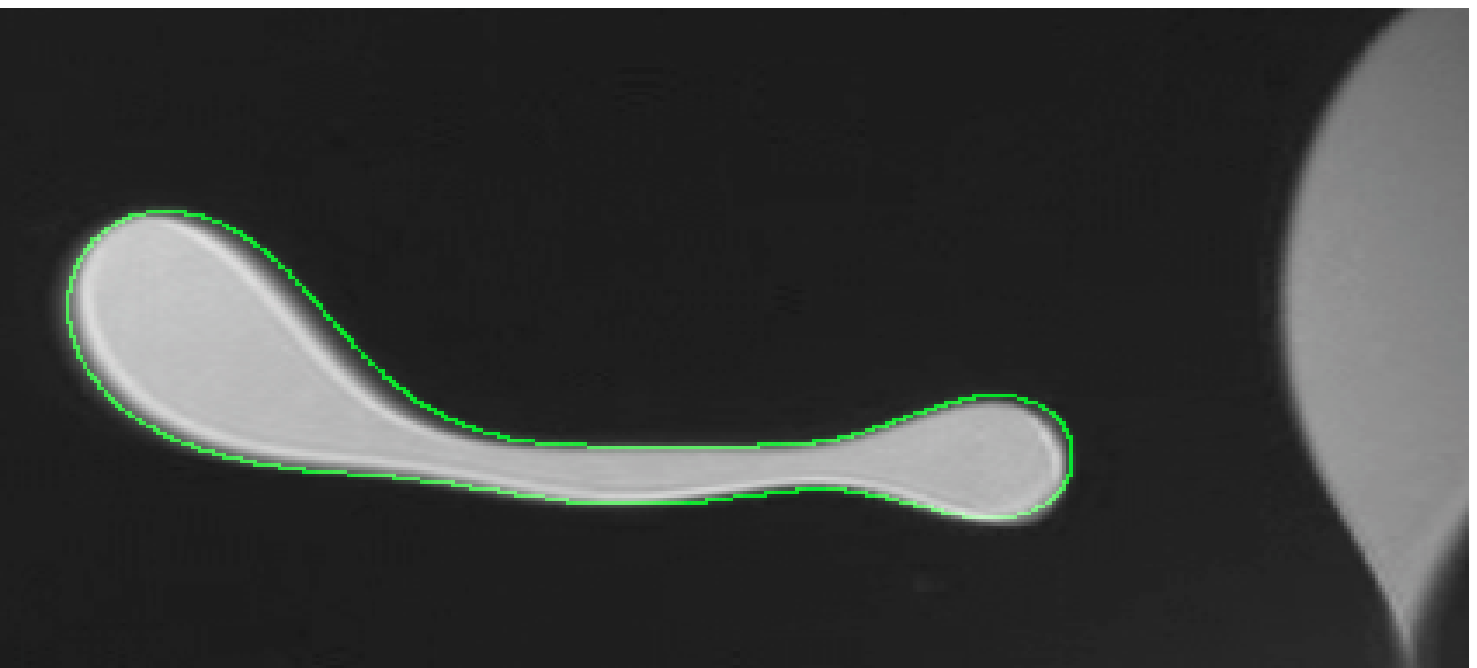} \\
\includegraphics[width= 80mm]{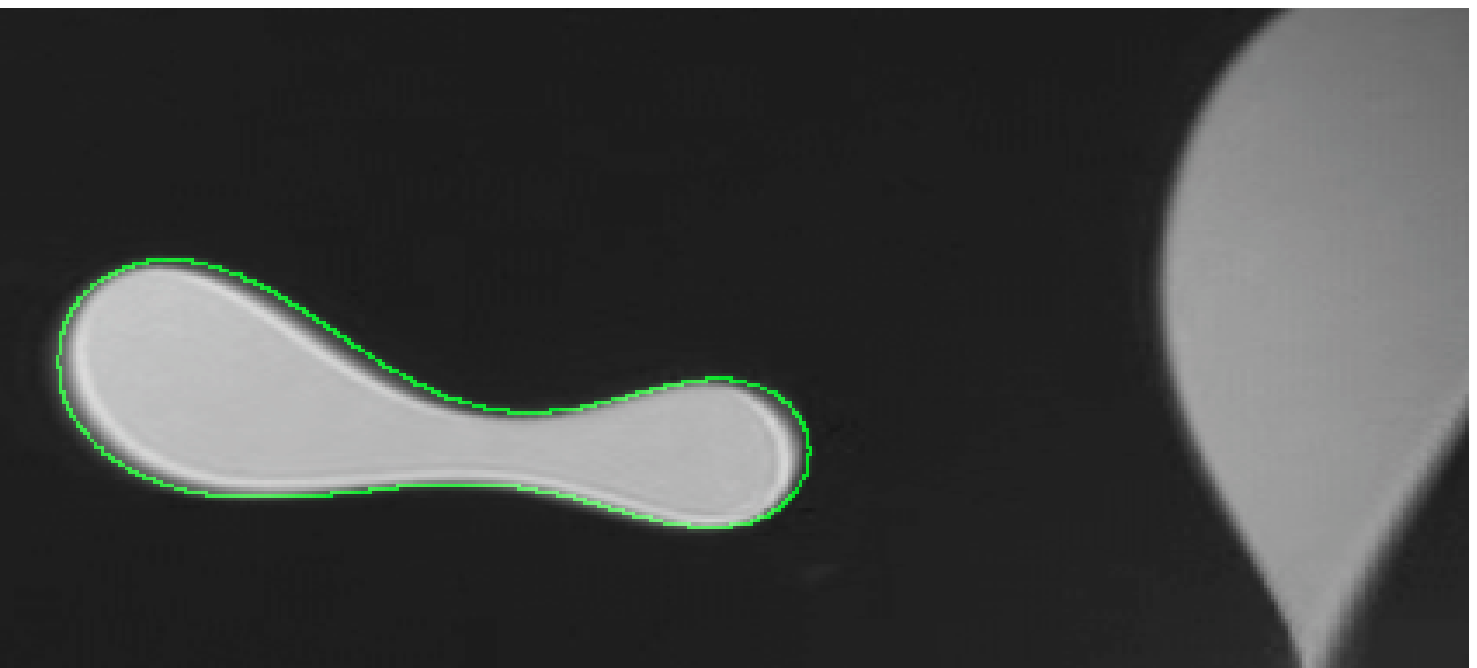}\\
\includegraphics[width= 80mm]{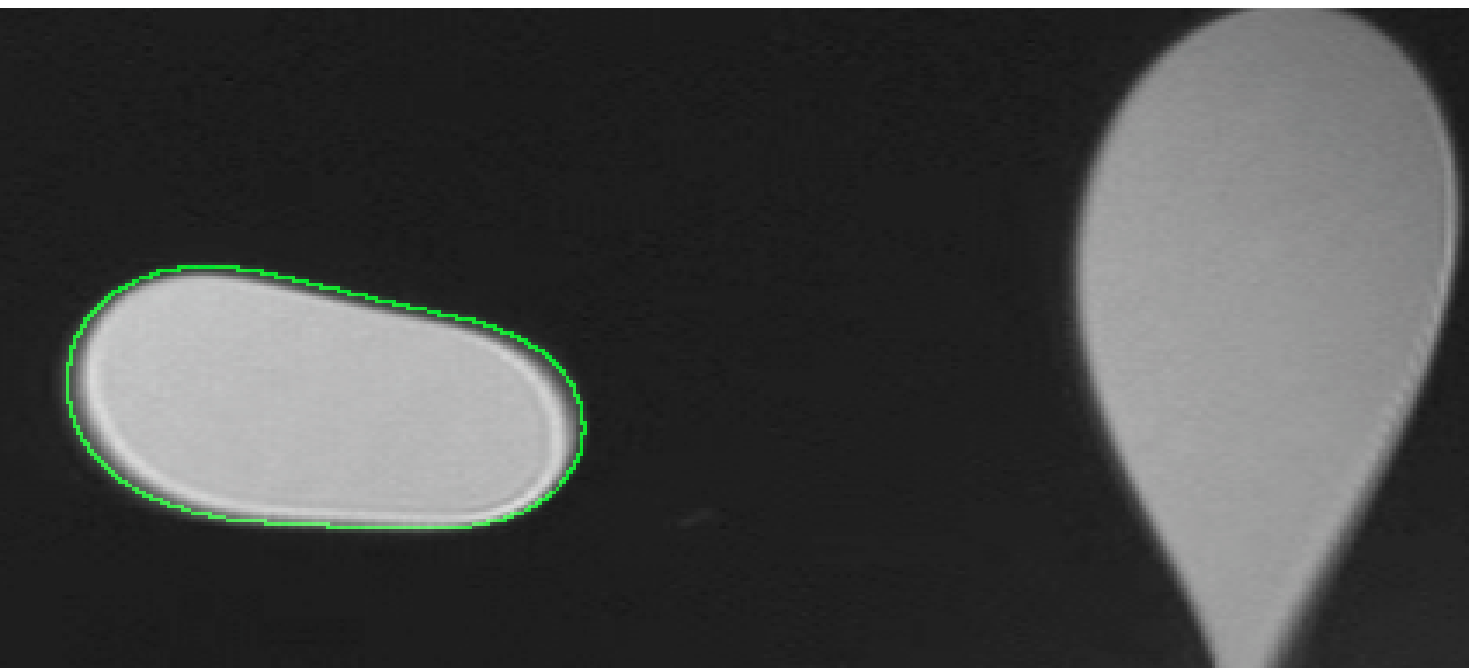}\\
\includegraphics[width= 80mm]{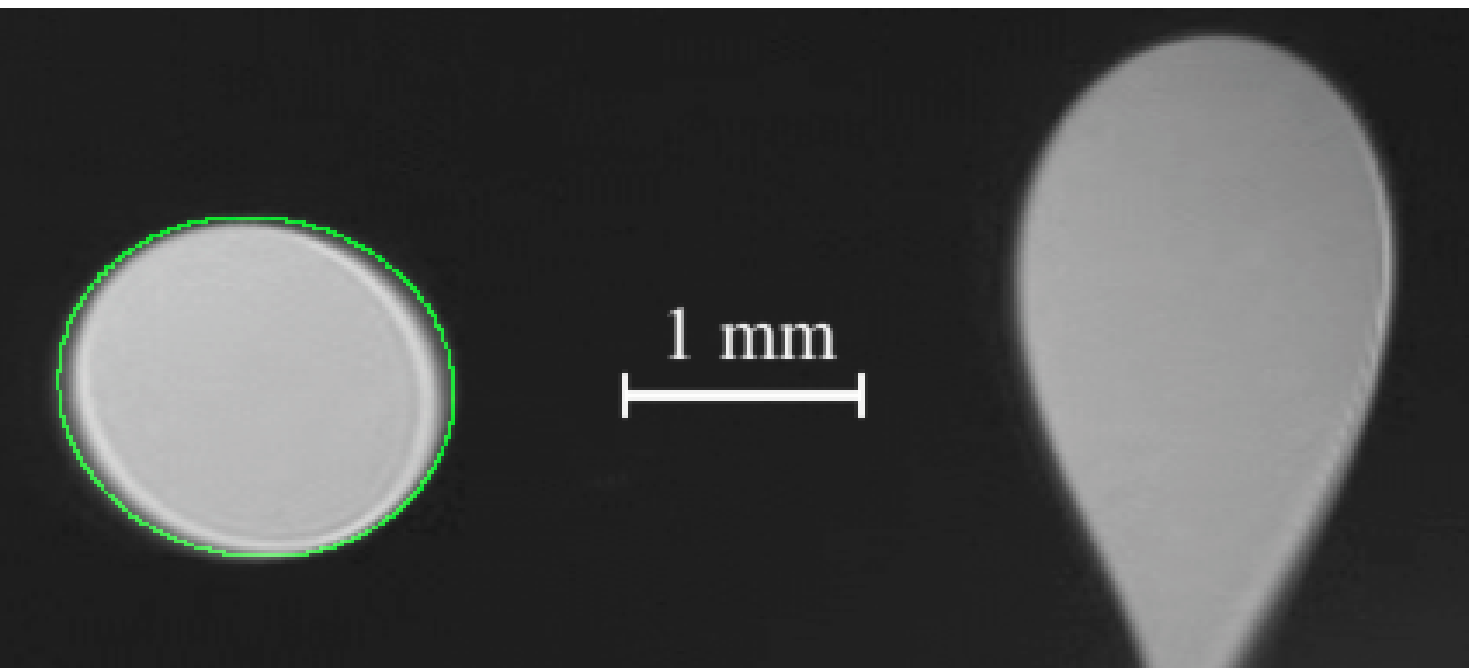}\\
\end{tabular}
\caption{ \label{fig:twist} A series of snapshots (for data set B)  showing the
relaxation of a stretched domain to a circle. The snapshots are separated by 2.85 seconds $\approx$ 0.558 $T_*$. The first photo is
superimposed with a curve marking the boundary determined by
our edge-finding algorithm; this is the initial condition for the
simulation of the domain relaxation. In subsequent photos the
superimposed curve shows the simulated shape of the bola, based on that 
initial condition. The simulation accurately
reproduces the observed dynamics even for complex, 
asymmetrical domains. The line
tension of this domain is found to be $390 \pm 3$ pN, where the uncertainty corresponds to the standard deviation of $T_*$. A movie of this relaxation is available from \texttt{http://www.math.hmc.edu/\~{}ajb/bola/simulation.mpeg}.}
\end{figure}

\section{Determining $\lambda$ via Experiment/Simulation Comparison }

Our equation of motion (\ref{eq:EOM})  is written in time units of  the
characteristic relaxation time, defined by  $T_* = \frac{\eta'
A_*}{\lambda}$. We calculate the line tension, $\lambda$, given
these other values. The domain's area, $A_*$,   is determined from
the photograph once the boundary curve is found, and the subfluid
viscosity, $\eta'$, is estimated from tabulated values (adjusting for
temperature). We determine $T_*$ by simulating the evolution and
matching timescales between the observed relaxation (snapshot times are recorded in seconds) and the simulated relaxation
(done in units of $T_*$).

We choose an initial condition taken from one of the snapshots and
simulate the subsequent relaxation.  For each photograph after this
first one, we match timescales by finding the time in the simulation
when the shape of the simulated domain most closely matches the domain shape
in the photo. We search through
the discrete time-steps of the simulation and compare the shape at
each step to the shape taken from the photo. Each snapshot gives us a value for $T_*$, computed as
\begin{equation}
T_* = \frac{t_j - t_0}{T_{best}-0}
\end{equation}
where $t_j$ and $t_0$ are the observed times of the comparison
snapshot and initial-condition snapshot, respectively, and
$T_{best}$ is the time which elapses in the simulation between $T=0$
(the initial condition) and the time at which the best-matching
shape occurs.

\begin{figure}
\includegraphics[width=80mm]{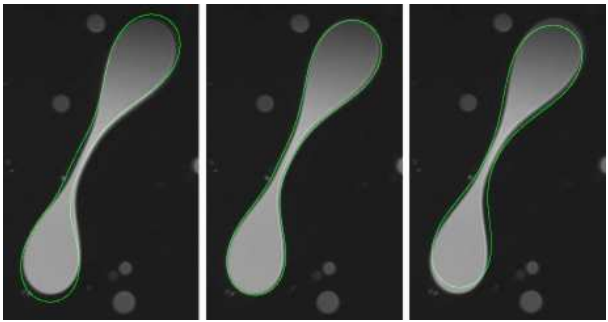}
\caption{ Three images showing the same photo overlayed
with simulated boundary curves from three different times in the
simulation. The center image shows the simulated curve at the time,
$T_{best}$, when it most closely matches the photo; the left and
right images show the simulated curve at $T = T_{best} - 0.1 T_*$
and $T = T_{best} + 0.1 T_*$, respectively. \label{fig:trio} }
\end{figure}

To measure how closely two domains (i.e. image-processed experiment and numerical simulation) match we use the Symmetric Difference Metric (SDM), which is determined by 
overlaying the domains and computing the total area which
lies in one  or the other but not both. For each photo, we
search through the time-steps of the simulation to find the
step at which the SDM between simulation domain and the experimental domain  is minimized; Figure \ref{fig:trio} provides a
visual illustration of this process. The minimum SDM over the
simulation provides a measure of how well each photo matches
\emph{some shape which occurs in the simulation};
we expect the same value for $T_*$ from every photo. We compute the mean value of the set of  $T_*$'s from
all the photos to determine the line tension, and the standard
deviation of these $T_*$'s provides an estimate for the precision in
the resulting measurement of $\lambda$.

To simulate the relaxation we must know the component of the
subfluid velocity which exists independent of (i.e. is not
directly produced by) the relaxation. Unfortunately, we cannot
directly measure the subfluid velocity. Instead, we choose an initial snapshot when the
subfluid is relatively quiescent and run the simulation under the
approximation that the ``independent'' subfluid velocity is zero.
We find, however, that the violation of this approximation is one of
the largest sources of systematic error contributing to mismatch
between observed and simulated domain shapes. We could choose a later
initial condition, waiting until remnant subfluid velocity is
negligible; however, this means throwing out a large portion of the
data (often all of it).

The type of motion which persists in the subfluid for the longest
time is solid-body motion; other types of motion are viscously
damped. We therefore correct for solid-body motion in the
post-simulation timescale fitting. Whenever we compare two shapes,
we do not directly compute the SDM between them, but instead determine the minimum SDM which can
be achieved by positioning one on top of the other using a
solid-body motion. This greatly reduces the SDM and allows
us to achieve excellent fits for data which would otherwise be
rendered worthless by remnant subfluid velocity.

Following \citep{Mann95}, we also measure the line tension by measuring the relaxation of small elliptical deformations of the boundary in the near circular limit, which we refer to as $\lambda_\epsilon$. The snapshots of the domain boundary are image processed,
and FFT techniques are used to extract the amplitude of the elliptical ($n=2$) deformation.  We then fit the exponential relaxation rate of this mode in the small amplitude limit. 

\begin{table}
\begin{tabular}{c|c|c|c|c|c|}
Data Set & $ \bl $  [pN] &	$\sigma_\lambda / \bl $ &	Average SDM &	$ \sigma_A/A_{avg}$  & $\lambda_{\epsilon}$ [pN] \\
\hline
A 
&	538 	 	& 3.4\%	&	3.9\%	&	1.0\%	& 468 \\	
B 
&	390   & 0.8\%	&	2.9\%	&	0.7\%			& 375		\\
C 
&	357  	&	1.7\%	&	3.0\%	&	0.3\%			& 362		\\
D 
&	570  	&	2.0\%	&	1.5\%	&	0.1\%			& 606		\\
E 
&	479  	&	4.0\%	&	1.5\%	&	1.0\%		& 485		\\
F 
&	191  	&	0.4\%	&	3.4\%	&	0.5\%			& 217 	 \\

\end{tabular}
\caption{ \label{tab:TheNumbers}  Line tension values and error estimates for six data sets. 
Here $\lambda_{avg}$ is the line tension values averaged over snapshots from large aspect ratio of bolas through the relaxation to nearly circular domains. The percentage error in this measurement is estimated by normalizing the standard deviation of these measurements by the average line tension, $\sigma_\lambda / \bl $. The average of the SDM (Symmetric Difference Metric) normalized by the domain area in these snapshots reflects how closely the numerical simulation reproduces the experimental results.
The standard deviation of the domain area, $\sigma_A$ divided by the average area, $A_{avg}$, 
is a proxy for the error in the image processing of the boundary. Finally, $\lambda_\epsilon$ is the line tension estimated from small amplitude perturbations from the final circular domain shape.}
\end{table}

Comparison data for six separate relaxations is presented in Table \ref{tab:TheNumbers}.
In six time series of different domains, the mean SDM between the experimental and simulated domains was 1.5-4\% of the domain's area, indicating that the proposed hydrodynamical model of  domain evolution reproduces the shapes quite well; this is clear from the comparison snapshots in the evolution in Figure \ref{fig:twist}.  The areas of the domains were constant across the time series to 0.1-1\%, well within the uncertainties of the measurement due
to diffraction at the domain edges. By matching time scales between the experiment and simulation, each photo after the first yields a value for $T_*$ in seconds; we deduce $\lambda_{avg}$ from these values. The percentage deviation of the values for $T_*$ from a set of photos ranges from 0.4\%-4\%, which also provide an error estimate on the line tension.

The greatest variances in $T_*$ and the largest SDM values (i.e. shape mismatches) occurred in those data sets where either (1) other domains nearly
touched the domain of interest or (2) remnant subfluid shearing was particularly problematic. Provided that reasonably isolated domains can be
produced and the subfluid flow can be well-controlled, it is possible to determine the line tension to a precision of $\sim \!\! 1\%$.

Finally, we note that the line tension estimates, $\lambda_\epsilon$,  from small perturbations of the final circular  shape are off by up to 13 \% -- this is comforting, in that it is consistent with
our results and variations observed in previous work \citep{Mann95}. It suggests that our new methodology is both accurate and precise.

\section{Conclusions}

In this paper we have described a method for determining the line tension driving the evolution of Langmuir layers. We are able to verify that our hydrodynamic model is consistent with the experiments and to determine the line tension with errors as small as 1\%, 
more than an order of magnitude better than previous efforts.

While we believe our measurement are accurate, it is striking that we have observed a wide variation in line tensions (191-570 pN) for the 8CB system. One factor that contributes significantly to this variation is the thickness of both the compact domain and its surroundings.  Experiments reported elsewhere 
\cite{Zou07}, using the relaxation of small deformations generated with a different deformation technique, systematically explore the dependence on the thickness of a compact domain 1-15 bilayers thick (on top of an unpaired monolayer) in a trilayer background.  These experiments lead to a line tension, reflecting the elastic energy of the dislocation at the domain boundary, proportional to the Burgers vector \cite{Zou07}. In the experiments reported here, we estimate, from the observed brightness of the domains, that the lighter compact domains range from 10-24 bilayers thick  and that the darker surrounding region is either three or five layers thick. Because of light scattering from the four-roll mill in our study the thickness of the dark region, expected to contribute to the line tension \cite{Geminard98}, is particularly uncertain. Other possible factors influencing the line tension include contamination and splitting of the boundary into two dislocation lines for very thick films \cite{Geminard98}.  Our expectation is that by quantifying this system and others more carefully we will be able to determine the dominant causes of line tension and generate reproducible results. This is a promising area for future investigation.

The present model assumes that bulk viscosity dominates the relaxation and that  both slip
between layers and electrostatic effects are negligible. These
conditions must be evaluated on a case-to-case basis. However, the
current method can determine the line tension with any technique
exploiting domain hydrodynamic response, including relaxation after
coalescence of two domains \citep{Roberts97} or after stretching a domain with
lasers tweezers \citep{Wurlitzer00}. It can also be generalized to more complex
situations involving three-phase contacts.

The boundary integral formulation here can be extended to incorporate more general potential forces such as electrostatics (cf. \citep{LubenskyG96,  HeinigHF04}) and as such we believe that we have developed a valuable tool for deducing and verifying the form of the intermolecular potential in systems that exhibit more complex morphology such as circle-to-dogbone transitions and labyrinth formation \citep{DeKokerM93, McConnell91}. This promises to be fertile ground for future research.

\section*{Acknowledgments}

A portion of this research was conducted by JRW and AJB as part of the UCLA Summer RTG program supported by NSF grant DMS-0601395, DMS-053552, and Harvey Mudd College Beckman and Presidential Research Grants.  LZ and EKM were partially supported by the NSF grant  DMR-9984304. JAM gratefully acknowledges the financial support from the MURI program,  ARO grant DAAD 19-03-1-0169. We wish to thank Julie Kim for purifying our 8CB and Prem Basnet  for fine-tuning the four-roll mill. Many of the numerical calculations were performed on the Harvey Mudd College Amber parallel cluster operated by the Mathematics and Computer Science departments.


\end{document}